\documentclass[a4, usenatbib]{mn2e}

\usepackage{epsfig}
\usepackage{amsmath}
\usepackage{natbib}

\def\apj{ApJ}

\def\apjs{ApJS}
\def\mnras{MNRAS}

\def\nat{Nature}
\def\prd{Phys.Rev D}

\begin{document}

\title[Galaxy Merger Timescales]{Modelling Galaxy Merger Timescales and Tidal Destruction}
\author[Simha \& Cole]
{Vimal Simha$^{1,2,3,4}$, Shaun Cole$^{4}$\\
$^1$University of the Western Cape, Bellville, Cape Town 7535, South Africa\\
$^2$South African Astronomical Observatories, Observatory, Cape Town 7925, South Africa\\
$^3$African Institute for Mathematical Sciences, Muizenberg, Cape Town 7945, South Africa\\
$^4$ Institute for Computational Cosmology, Department of Physics, Durham University, South Road, Durham DH1 3LE, United Kingdom\\
e-mail: vimalsimha@gmail.com\\
}

\maketitle

\begin{abstract}

We present a model for the dynamical evolution of subhaloes based on an approach combining numerical and analytical methods. Our method is based on tracking subhaloes in an N-body simulation up to the last point that it can be resolved, and applying an analytic prescription for its merger timescale that takes dynamical friction and tidal disruption into account. When applied to cosmological N-body simulations with mass resolutions that differ by two orders of magnitude, the technique produces halo occupation distributions that agree to within 3\%.

\end{abstract}

\section{Introduction}

The evolution of the Universe in the standard cold dark matter model is characterized by hierarchical structure formation where small haloes form first, and subsequently merge to form larger haloes. High resolution N-body simulations have indicated that massive haloes retain a substantial amount of substructure \citep{klypin99,moore99,springel01}, consisting of bound dark matter clumps orbiting within the potential of their host halo. Evidently, such subhaloes were themselves independent, self-contained haloes in the past, before merging with a more massive halo. If sufficiently massive, these subhaloes were sites of baryon dissipation and star formation in the past. There are many indications, from studies of the statistical properties of how galaxies and sub-structures populate haloes, that galaxies in groups and clusters are in fact the observational counterparts of subhaloes. For example, \cite{colin99} and \cite{kravtsov04} show that the auto correlation functions of substructures in high resoution N-body simulations are in good agreement with the observed auto correlation functions of galaxies. On the theory side, \cite{kravtsov04} find that the distribution of subhaloes in high resolution N-body simulations is similar to that of smoothed particle hydrodynamics (SPH) galaxies in \cite{berlind03} and \cite{zheng05}, and \citep{simha12} find good agreement between the properties of galaxies in their SPH simulation, and the properties of halos in their matched N-body simulation.  

Using N-body simulations to study galaxy formation requires a framework for populating dark matter halos with galaxies. One approach is to employ semi-analytic models which use analytic techniques and phenomenological recipes to follow the evolution of baryons within dark matter halos. An alternative is to eschew assumptions about baryonic physics and use an empirical approach to relate observed galaxy properties to dark matter halo properties. One popular approach in this class of models is abundance matching which assumes a monotonic relationship between galaxy luminosity and halo mass.

A key aspect of studying galaxy formation within the cold dark matter paradigm is to understand the fate of galaxies following halo mergers - how and when galaxy mergers happen, how and when galaxies are tidally disrupted and what ultimately happens to galaxies that fall in to more massive haloes. In this paper, we model the dynamical evolution of subhaloes and their galaxies using a combination of numerical and analytical methods. 

The formation of dark matter haloes through the growth of dark matter perturbations can be studied numerically using dissipationless cosmological simulations. But haloes do not evolve in isolation from each other. When a halo enters the virial radius of a more massive halo, its evolution becomes more complex than that of independent haloes. After falling in to a more massive halo, (sub)haloes experience tidal forces which cause mass loss, and even complete disruption under extreme circumstances. Furthermore, satellite subhaloes orbiting within a host halo lose energy and angular momentum through dynamical friction which causes their orbits to sink towards the halo centre. 

Galaxy formation models use two principal approaches to model the dynamical evolution of subhaloes and satellite galaxies. The first is to use analytically calculated or physically motivated empirical formulae for various aspects of the dynamical evolution of subhaloes such as the timescale for merging through dynamical friction, tidal stripping, tidal destruction, etc. \citep[e.g.][]{lc,bk08,jiang08}. An alternative approach, that avoids simplifying assumptions, is to use an N-body simulation to follow the dynamic evolution of subhaloes. N-body simulations capture the complexity of the physics of tidal disruption and dynamical friction, and do not require simplifying assumptions, either about the physical processes or the distribution of orbital parameters. However, subhalo merging and disruption are affected by finite force and mass resolution. Insufficiently resolved subhaloes disrupt artificially and on shorter timescales than well resolved haloes \citep{klypin99}. Using a fixed subhalo mass resolution limit, regardless of infall mass, leads to lower mass subhaloes being artificially disrupted more quickly.  

In subhalo abundance matching (SHAM) models, which assume a monotonic relationship between subhalo mass at infall and galaxy luminosity, it is assumed that each galaxy survives as long as its subhalo can be identified in an N-body simulation above a fixed resolution threshold. Some models allow galaxies to survive for a period of time after the disruption of their host subhaloes \citep{saro08,moster10}, while in contrast, \cite{stewart09} allows satellite galaxies to be disrupted even while their subhaloes still exist.

In contrast, the GALFORM semi-analytic model uses an analytical formula to calculate the merger timescale for a satellite galaxy \citep{cole00}. Following a halo merger, it is assumed that the galaxy hosted by the less massive halo enters the host halo on an orbit with orbital parameters randomly drawn from a distribution. The merger timescale is then computed using the analytical formulae of \cite{lc} in the GALFORM model of \cite{gon14} and the analytical formulae of \cite{jiang08} in the GALFORM model of \cite{lacey15}. Once this time has elapsed, the galaxy hosted by this subhalo is considered to have merged with the central galaxy of the more massive host halo.

In this paper, we employ a hybrid approach to follow the dynamical evolution of subhaloes. We follow subhaloes in an N-body simulation until the point when they can no longer be resolved. We then calculate a merger timescale using its orbital parameters and mass at the epoch that it was last resolved in the N-body simulation. Our formula to calculate the merger timescale is based on \cite{lc}, with parameters suitably modified to match our N-body simulation results. Our scheme is faithful to the underlying N-body simulation, minimising the reliance on analytically determined orbits. Instead of making assumptions about the orbital parameters of satellites, we track the positions of their associated subhaloes. 

Our goal is to provide a simple model for the dynamical evolution of subhaloes that uses the information in an N-body simulation, but can produce results that are not affected by artificial disruption of subhaloes due to limited resolution. While our model is primarily intended for application in semi-analytic models, it can also be used in other models of galaxy formation that use N-body simulations like SHAM models. We assume that once a subhalo reaches the centre of its host halo, the galaxy associated with the subhalo merges with the central galaxy of the host halo. We also assume that once a subhalo is tidally disrupted, the galaxy it hosts is also tidally disrupted. Therefore, within the context of this paper, subhalo mergers and tidal disruption are synonymous with galaxy mergers and galaxy tidal disruption.

In \S2, we describe our simulation and models for populating our subhaloes with galaxies. In \S3, we describe our model for dynamical friction and tidal disruption. In \S4, we discuss our results. In \S5, we discuss the implications of our model on halo occupation distributions and galaxy clustering. In \S6, we summarise our results.

\section{Simulations}

We use two simulations, the Millennium Simulation \citep{springel05} and the higher resolution Millennium II 
simulation \citep{bk09}. Both simulations follow the evolution of 2160$^3$ particles from 
$z=127$ to $z=0$ in a $\Lambda$CDM cosmology (inflationary, cold dark matter with a 
cosmological constant) with $\Omega_{\rm M}$=0.25, $\Omega_{\Lambda}$=0.75,
$h\equiv H_0/100$ km s$^{-1}$Mpc$^{-1}$= 0.73, 
primordial spectral index $n_s$=1, and the amplitude of mass fluctuations, $\sigma_8$=0.9 where $\sigma_8$ is 
 the linear theory rms mass fluctuation amplitude in spheres of radius 8 $h^{-1}$Mpc at $z=0$. These parameter values 
 were chosen to agree with WMAP-1yr data \citep{spergel03}, and are 
different from, but reasonably close to current estimates from the cosmic microwave background
\citep{planck15}. The main difference is that the more recent data favour a lower 
value of the amplitude of clustering, $\sigma_8$. We do not expect small differences in the cosmological parameters to affect our results.

The Millennium Simulation (MS) simulates a comoving box that is 500$h^{-1}$Mpc on each side while the Millennium II Simulation (MS-II) simulates a comoving box that is 100$h^{-1}$Mpc on each side. The simulation particle masses are $m_{\rm p}$ = 8.6$\times$10$^8$ $h^{-1}{\rm M}_{\odot}$ in MS and 6.9$\times$10$^6$ $h^{-1}{\rm M}_{\odot}$ in MS-II. 

For each output epoch in each simulation, the friends-of-friends (FOF) algorithm is used to identify groups by linking together particles separated by less than 0.2 times the mean inter particle separation \citep{davis85}. The SUBFIND algorithm \citep{springel01} is then applied to each FOF group to split it into a set of self-bound subhaloes. The central subhalo is defined as the most massive subunit of a FOF group. We construct subhalo merger trees which link each subhalo at each epoch to a unique descendent in the following epoch. These merger trees allow us to track the formation history of each (sub)halo that is identified at $z=0$. \cite{springel01} and \cite{bk09} provide a detailed description of these simulations and the post-processing techniques.

\subsection{SHAM}

Subhalo abundance matching (SHAM) is a technique for assigning galaxies to simulated dark matter haloes and subhaloes. The essential assumptions are that all galaxies reside in identifiable dark matter substructures and that luminosity or stellar mass of a galaxy is monotonically related to the potential well depth of its host halo or subhalo. Some implementations use the maximum of the circular velocity profile as the indicator of potential well depth, while others use halo or subhalo mass. The first clear formulations of SHAM as a systematic method appear in \cite{conroy06} and \cite{vale06}, but these build on a number of previous studies that either test the underpinnings of SHAM or implicitly assume SHAM-like galaxy assignment \citep[e.g.][]{colin99,kravtsov04,nagai05}.

N-body simulations produce subhaloes that are located within the virial radius of haloes. The present mass of subhaloes is a product of mass build up during the period when the halo evolves in isolation and tidal mass loss after it enters the virial radius of a more massive halo \citep[e.g.][]{kravtsov04,kazantzidis04}. The stellar component, however, is at the bottom of the potential well and more tightly bound making it less likely to be affected by tidal forces. Therefore, several authors \citep[e.g.][]{conroy06,vale06} argue that the properties of the stellar component should be more strongly correlated with the subhalo mass at the epoch of accretion rather than at $z=0$.

\cite{vale06} apply a global statistical correction to subhalo masses relative to halo masses (as do \citealt{weinberg08}), while \cite{conroy06} explicitly identify subhaloes at the epoch of accretion and use the maximum circular velocity at that epoch. Our formulation here is similar to that of \cite{conroy06}, though we use mass rather than circular velocity. Specifically, we assume a monotonic relationship between galaxy luminosity and halo mass (at infall) and determine the form of this relation by solving the implicit equation
\begin{equation}
 n_S(>M_r) = n_H(>M_H),
\end{equation}
where $n_S$ and $n_H$ are the number densities of galaxies and haloes, respectively, $M_r$ is the galaxy $r$-band magnitude threshold, and $M_H$ is the halo mass threshold chosen so that the number density of haloes above it is equal to the number density of galaxies in the sample.
The quantity $M_{\rm H}$ is defined as follows:
\begin{equation}
M_H=\begin{cases}
M_{\rm halo} (z=0) & \text{for distinct haloes},\\
M_{\rm halo} (z=z_{\rm sat})& \text{for subhaloes},
\end{cases}
\end{equation}
where z$_{\rm sat}$ is the epoch when a halo first enters the virial radius of a more massive halo.

\subsection{GALFORM}
\label{galform}

GALFORM is a semi-analytic model of galaxy formation and evolution. In GALFORM-GP14 \citep{gon14}, the timescale for merging satellite galaxies with the central galaxy of its host halo due to dynamical friction is calculated according to \cite{cole00} and \cite{lc}. In this model, following a halo merger, each subhalo, along with the satellite galaxy it contains, is placed on a random orbit. A merger time scale is then calculated using equation \ref{lceq}. The treatment of galaxy mergers in \cite{lacey15} is similar to \cite{gon14}, except that they replace the dynamical friction timescale formula of \cite{lc} with that of \cite{jiang08} which implicity accounts for the effect on the dynamical friction timescale caused by tidal stripping. The satellite galaxy is considered to have merged with its central galaxy once the merger time-scale has elapsed, provided that this happens before its host halo falls in to an even larger system, in which case a new merger time-scale is computed. In this paper, we do not make use of any of the baryonic physics implemented in GALFORM.

\section{Model}

Following halo mergers, the less massive halo continues to exist as a distinct subhalo within the more massive host halo for a period of time. Its independent existence can come to an end in one of two distinct ways. Firstly, subhaloes can merge with the central halo after being drawn in to the centre of the host halo by dynamical friction. Secondly, subhaloes experience tidal stripping, and if sufficiently close to the centre of the host halo, they can be completely disrupted tidally. In this section, we outline how we model each of these processes.

\subsection{Dynamical Friction}

Subhaloes in orbit within a more massive host halo experience forces from the particles of the host halo which dissipate its energy and angular momentum, and drag it towards the centre of the host halo, where the galaxy it hosts can merge with the central galaxy of the host halo. 

Starting from the calculation of acceleration from dynamical friction by \cite{chandra43}, \cite{lc} derive the following expression for the merger timescale, $T_{\rm df}$, for subhaloes entering the virial radius of a more massive host halo with a singular isothermal sphere density profile.
\begin{equation}
T_{\rm df} = \frac{f(\epsilon)}{2B(1){\rm ln}{\Lambda}} \left(\frac{R_{\rm H}}{V_{\rm c}}\right) \left(\frac{r_{\rm c}}{R_{\rm H}}\right)^2 \left(\frac{M_{\rm H}}{M_{\rm S}}\right)
\label{lceq}
\end{equation}
where $\epsilon$ = $J/J_{\rm c}$, is the ratio of the angular momentum of the actual orbit to the angular momentum of a circular orbit with the same energy, $r_{\rm c}$ is the radius of a circular orbit in the halo with the same energy as the actual orbit, $r$ is the radius of the actual orbit, $R_{\rm H}$/$V_{\rm c}$ = $\tau_{\rm {dyn}}$ is the dynamical time of the halo, $M_{\rm H}$ and $M_{\rm S}$ are the masses of the host halo and subhalo respectively, ln$\Lambda$ is the Coulomb logarithm taken to be ln$(M_{\rm H}/M_{\rm S})$ and
\begin{equation}
B(x) = {\rm erf}(x) - \frac{2{x}}{\sqrt\pi} exp(-{x}^2).
\end{equation} 

Although, equation \ref{lceq} captures the essential characteristic features of the merging of subhaloes due to dynamical friction, it makes certain simplifying assumptions. Firstly, the density profile of dark matter haloes is assumed to be a singular isothermal sphere. While halo density profiles are approximately isothermal over a large range in radius, N-body simulations have found that they are significantly shallower than $r^{-2}$ at small radii and steeper than $r^{-2}$ near the virial radius \citep{nfw96}. Halo density profiles are better approximated by an NFW profile given by:
\begin{equation}
\rho(r) = \frac{\rho_0}{(r/r_s){(1+r/r_s)}^2}
\end{equation}
where $r_{\rm s}$ = $r_{200}$/$c$, $r_{200}$ is the radius at which the average density of the halo is 200 times the critical density and $c$ is the halo concentration parameter. Additionally, contrary to the assumptions made to carry out the above calculation, the host halo is non-spherical, its velocity dispersion is not necessarily isotropic and it is usually evolving. 

In our model, following each halo merger, we track each subhalo until the point that it can no longer be resolved in the N-body simulation. We then compute the energy of the subhalo at the latest epoch that it was identified in the N-body simulation, assuming the host halo to have an NFW density profile with concentration parameter, $c$ given by the halo mass-concentration relation. We then determine the dynamical friction timescale, $T_{{\rm df}}$ from equation \ref{oureq} below:
\begin{equation}
T_{\rm df} = \left(\frac{R_{\rm H}}{R_{\rm c}}\right)^{\alpha} \left(\frac{J}{J_{\rm c}}\right)^{\beta} \frac{\tau_{\rm dyn}}{2B(1){\rm ln}{\Lambda}} \left(\frac{M_{\rm H}}{M_{\rm S}}\right)
\label{oureq}
\end{equation}

Equation \ref{oureq} is similar to equation \ref{lceq} of \cite{lc}, but with the terms rearranged for convenience. \cite{lc} set $\alpha$=2 and $\beta$ = 0.78. Their value of $\alpha$ comes from the analytic calculation assuming a singular isothermal sphere density profile for the host halo and their value of $\beta$ was determined from numerical integration of the orbit-averaged equations for energy and angular momentum loss due to dynamical friction for a point mass in a singular isothermal sphere potential. In contrast to \cite{lc}, we treat the dependence of the dynamical friction timescale on energy and angular momentum as free parameters. We determine both $\alpha$ and $\beta$ numerically, finding $\alpha$ = -1.8 and $\beta$ = 0.85. We discuss the details of the procedure for determining these parameters in \S4.

If the parent halo merges with an even larger halo before time $T_{\rm df}$, we recalculate a new dynamical friction time scale, $T_{{\rm df}}$ for the new host halo. We assume that the subhalo retains its mass and ignore the effect of tidal stripping. We ignore interactions between two orbiting subhaloes. Such interactions rarely result in mergers because satellite subhaloes are unlikely to encounter other satellite subhaloes at low enough velocities to result in a bound interaction \citep{wetzel09}. 

\subsection{Tidal Disruption}

A subhalo in orbit within a more massive host halo experiences tidal forces which can strip away the outer regions, or, in some cases, entirely disrupt the subhalo. Material is stripped from satellite subhaloes when the tidal force from the host halo exceeds the self-gravity of the subhalo, and the same process under extreme circumstances leads to complete disruption of the satellite subhalo.

In our model, we disrupt a satellite subhalo if the mean density of the host halo within the radius of the satellite subhalo exceeds the density of the satellite subhalo i.e. if the distance of the subhalo from the centre of the host halo falls below a tidal disruption radius, $R_{\rm {td}}$, which is defined as the radius within which the mean density of the host halo exceeds the density of the satellite subhalo.

\begin{equation}
R_{\rm {td}} = r_{\rm{sat}} \sqrt[3]{\frac {M_{\rm H}(<{R_{\rm {td}}})}{M_{\rm{sat}}}},
\label{otd}
\end{equation}

where $R_{\rm td}$ is the tidal disruption radius, $r_{\rm{sat}}$ is the radius of the satellite subhalo, $M(<{R_{\rm td}})$ is the mass of the host halo enclosed within $R_{\rm td}$ and $M_{\rm{sat}}$ is the mass of the satellite subhalo. The mass of the host halo enclosed within $R_{\rm td}$ is determined assuming an NFW density profile for the host halo with concentration parameter $c$ determined from the halo mass-concentration relation, but alternatively it can be measured in the simulation.
 
We follow the position of a subhalo as long as it exists in the N-body simulation. We then compute the position of its most bound particle at the latest epoch that the subhalo was resolved. We then follow the position of this particle and treat it as the position of the subhalo. The subhalo is removed from the population if its distance from the centre of the host halo falls below the tidal disruption radius, $R_{\rm {td}}$.

We ignore mass loss due to tidal stripping. In our model, tidal destruction is treated as a binary process that takes effect and removes a subhalo and the galaxy it hosts from the population only when the subhalo enters a region that is denser than its own mean density. 

\section{Results}

\begin{figure}
\centerline{
\epsfxsize=84mm
\epsfbox{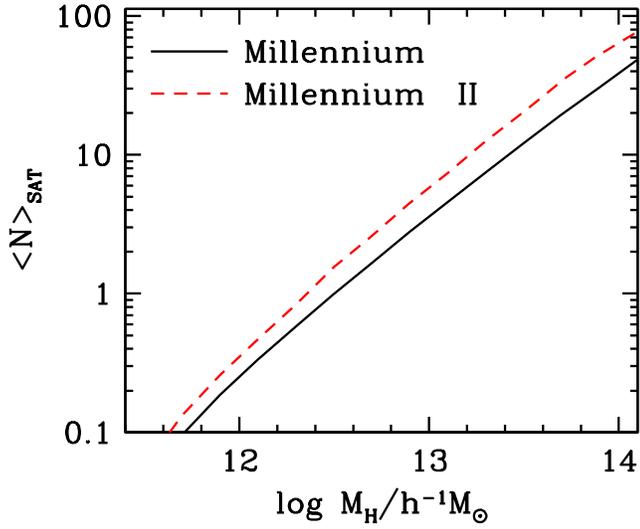}
}
\caption{
Mean number of subhaloes as a function of parent halo 
mass at $z=0$ in Millennium (black solid) and Millennium II (red dashed) simulations. Subhaloes above an infall mass threshold of 8.6$\times$10$^{10}$ 
$h^{-1}$M$_{\odot}$ are selected which 
corresponds to 100 times the particle mass in the Millennium Simulation.
}
\label{fig:sub1}
\end{figure}

Fig. \ref{fig:sub1} compares the mean number of subhaloes as a function of halo mass between the Millennium (MS) and Millennium II (MS II) simulations, two N-body simulations whose mass resolutions differ by a factor of 100. We select subhaloes that are above an infall mass threshold of 8.6$\times$10$^{10}$ $h^{-1}$M$_{\odot}$ which corresponds to 100 times the particle mass in MS. Although the halo mass function converges for this mass threshold \citep{bk09}, there are substantial differences in the subhalo population between MS and MS II. The higher resolution MS II retains substantially more subhaloes at $z=0$. Although MS can adequately resolve subhaloes with more than 100 particles at infall, subhaloes are subject to tidal stripping after infall which reduces their mass and thus renders them unresolvable in MS at later times including at $z=0$. However, similar mass objects can be resolved in MS II as they contain 100 times more particles. 

To track subhaloes that can no longer be resolved in the N-body simulation, we implement the procedure outlined in \S3 to subhaloes in both MS and MS II. In short, subhaloes that can no longer be resolved in the N-body simulation are removed from the population after time $T_{\rm df}$ given by equation \ref{oureq} has elapsed from the last epoch the subhalo was resolved in or if the pericentric distance between the subhalo and the centre of the host halo falls below $R_{\rm {td}}$ given by equation \ref{otd}. 

Fig. \ref{fig:sub6} shows the distribution of energy and angular momentum for haloes that merged with a more massive halo between $z=1$ and $z=0$. It includes subhaloes that merged with the parent halo between these epochs, and those that continue to exist at $z=0$. More than half of all subhaloes have $r/r_{\rm c}$ greater than 1. Most subhaloes have $J/J_{\rm c}$ between 0.6 and 0.9.

Fig. \ref{fig:sub7} shows the distribution of mass ratios for the same sample. Nearly 90\% of subhaloes have less than one-fifth of the mass of the host halo at infall. The number of nearly equal mass mergers is small, and in any case, most merge with the host halo on fairly short timescales.

Fig. \ref{fig:sub2} shows the mean number of subhaloes as a function of halo mass upon applying our model to MS and MS II. Using the canonical values of $\alpha$ = -2 and $\beta$ = 0.78 in equation \ref{oureq} produces substantially better agreement between the mean number of subhaloes as a function of host halo mass between MS and MS II compared to Fig. \ref{fig:sub1}. However, it can be further improved by fitting our model parameters, $\alpha$ and $\beta$ to minimise the difference in the number of subhaloes in bins of host halo mass between MS and MS II. We fit $\alpha$ and $\beta$ by minimising the quantity $| \Delta \langle N_{\rm sub} \rangle|$ in bins of halo mass, finding $\alpha$ = -1.8 and $\beta$ = 0.85. 

We have checked that our results are not sensitive to the mass threshold used in Fig.s  \ref{fig:sub1} and \ref{fig:sub2}. 

\section{Implications for Galaxy Clustering}

In this section, we investigate the implications of our model for the dynamical evolution of subhaloes for studies of galaxy clustering. One plausible method of comparing our model to the raw N-body simulation would be to apply identical baryonic physics to the raw N-body merger trees and those derived by applying our model to the N-body merger trees. However, such an approach would produce galaxy catalogues with different galaxy stellar mass functions. An alternative approach is to tune the baryonic physics independently in each case to match certain observables like the galaxy stellar mass function. However, that would still leave us with the difficulty of disentangling differences due to different baryonic physics from the differences due to the treatment of the dynamic evolution of subhaloes.

In this paper, instead of explicit assumptions about the baryonic physics, we employ subhalo abundance matching to construct our galaxy catalogues. We compare our model for the dynamical evolution of subhaloes with two other models for the dynamical evolution of subhaloes which produce different populations of surviving subhaloes. Firstly, subhalo abundance matching (SHAM) carried out on the raw N-body simulation where we consider only subhaloes that survive until $z=0$ in the N-body simulation. Secondly, the method used by GALFORM-GP14 for evolving subhaloes and satellite galaxies discussed in \S\ref{galform}. We do not make use of any of the baryonic physics implemented in GALFORM. 

In all three cases, we populate our subhaloes with galaxies using subhalo abundance matching. The number density of objects is 3.03$\times$10$^{-2}h^{3}$Mpc$^3$ which corresponds to the number density of objects brighter than $M_{\rm r}$ of -18 in SDSS DR7 \citep{zehavi10}. In all three cases, the number density of galaxies and the galaxy luminosity functions are identical. The differences between the catalogues arise solely from the different treatment of subhaloes following halo mergers which manifest themselves as differences in the satellite galaxy fractions which in turn affect halo occupation distributions and clustering statistics.

Fig. \ref{fig:sub3} compares the $z=0$ halo occupation distribution of our model to that obtained from abundance matching on the raw N-body merger trees and GALFORM-GP14's treatment of the evolution of subhaloes. At the resolution of Fig. \ref{fig:sub3}, where subhaloes have $\approx$2$\times$10$^4$ particles, the differences between our model and the raw N-body merger trees are small, although our model will recover a similar halo occupation distribution even with a lower resolution N-body simulation. However, the GALFORM-GP14 model retains substantially more subhaloes than either of the other two models. Note that all models are constrained to have the same number of galaxies, but each model has a different number of central and satellite galaxies. While the satellite fractions of our model and SHAM on the raw N-body merger tree are within 1\% of each other, the satellite fraction in the GALFORM-GP14 model is 19\% higher.

Fig. \ref{fig:sub4} compares the $z=0$ projected two-point correlation function of galaxies in our model to that obtained from abundance matching on the raw N-body merger trees and GALFORM-GP14's treatment of the evolution of subhaloes, and Fig. \ref{fig:sub5} shows the ratio of the correlation function produced by abundance matching on the raw N-body merger trees and GALFORM-GP14's treatment of the evolution of subhaloes and our model. On scales below $\approx$1$h^{-1}$Mpc, the galaxy two-point correlation function is determined by pairs of objects within haloes - the `one-halo term'. Since the GALFORM-GP14 model contains 19\% more satellite galaxies, it produces a higher clustering amplitude which can be up to twice as high as the clustering amplitude in our model on certain scales. The clustering of galaxies in our model and abundance matching on the raw N-body merger trees agree to better than 5\%, reflecting the fact that they contain a similar fraction of subhaloes.

On scales larger than $\approx$1$h^{-1}$Mpc, the two-point correlation function is determined by pairs of galaxies in different haloes known as the `two-halo term'. Since the overall number density of objects and the positions of haloes are the same in all models, there are only small differences on these scales.

\section{Discussion and Conclusions}

We present a model for the dynamical evolution of subhaloes based on an approach that combines numerical and analytical methods. Our method is based on tracking subhaloes in an N-body simulation up to the point that it can be resolved, and applying an analytic prescription for its subsequent merger timescale that takes dynamical friction and tidal disruption into account. When applied to cosmological N-body simulations with mass resolutions that differ by two orders of magnitude, the technique presented in this paper produces halo occupation distributions that agree to within 3\%.

Modelling galaxy mergers within dark matter haloes is an important ingredient of galaxy formation models. Precise estimates of galaxy merger timescales are required for modelling galaxy clustering, mass assembly of galaxies, properties of satellite galaxies and black hole merger rates. 

While subhaloes that approach the host halo too closely can be tidally destroyed in our model, we do not model mass loss due to tidal stripping and its effects on the dynamical friction timescale. We also ignore satellite-satellite interactions which, in any case, are rare.

\begin{figure}
\centerline{
\epsfxsize=84mm
\epsfbox{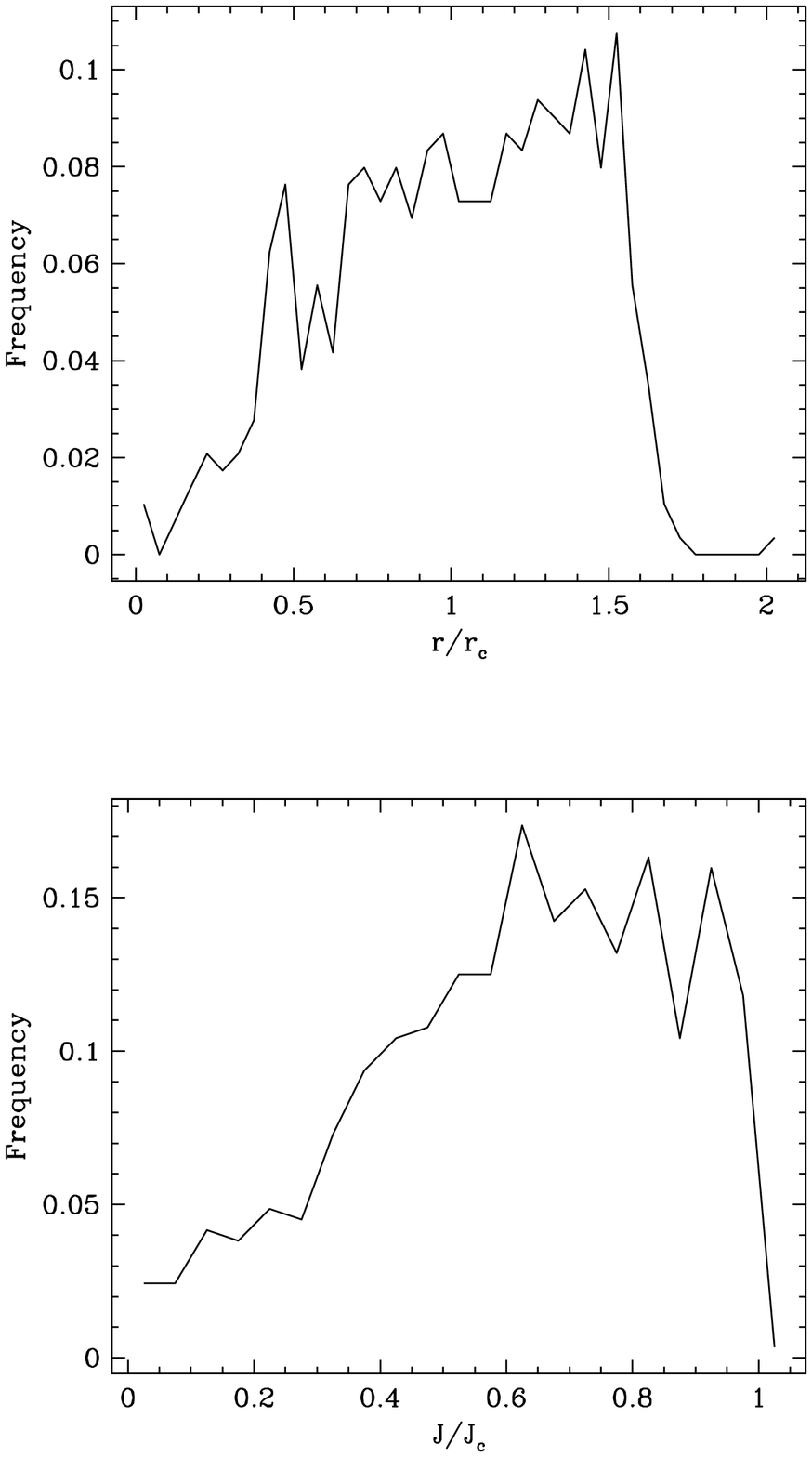}
}
\caption{
(Top) The distribution of $r/r_{\rm c}$ of all infalling objects between $z=1$ and $z=0$.
(Bottom) The distribution of $J/J_{\rm c}$ of all infalling objects between $z=1$ and $z=0$.
}
\label{fig:sub6}
\end{figure}

\begin{figure}
\centerline{
\epsfxsize=84mm
\epsfbox{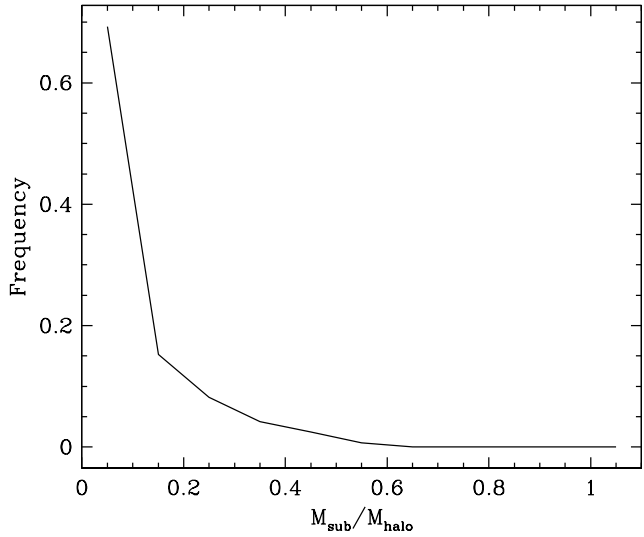}
}
\caption{
The distribution of mass ratios of all infalling objects between $z=1$ and $z=0$.
}
\label{fig:sub7}
\end{figure}

\begin{figure}
\centerline{
\epsfxsize=84mm
\epsfbox{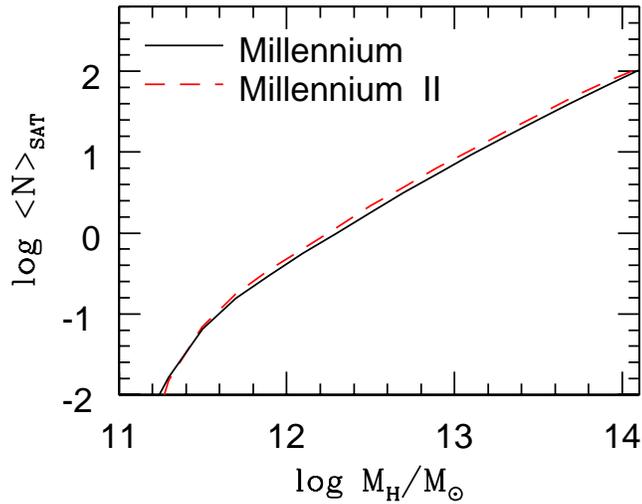}
}
\caption{
Mean number of subhaloes per halo  as a function of parent halo 
mass at $z=0$ after applying our model for the dynamical evolution of subhaloes. Subhaloes above an infall mass threshold of 8.6$\times$10$^{10}$ 
$h^{-1}$M$_{\odot}$ are selected which 
corresponds to 100 times the particle mass in the Millennium Simulation.
}
\label{fig:sub2}
\end{figure}

\begin{figure}
\centerline{
\epsfxsize=84mm
\epsfbox{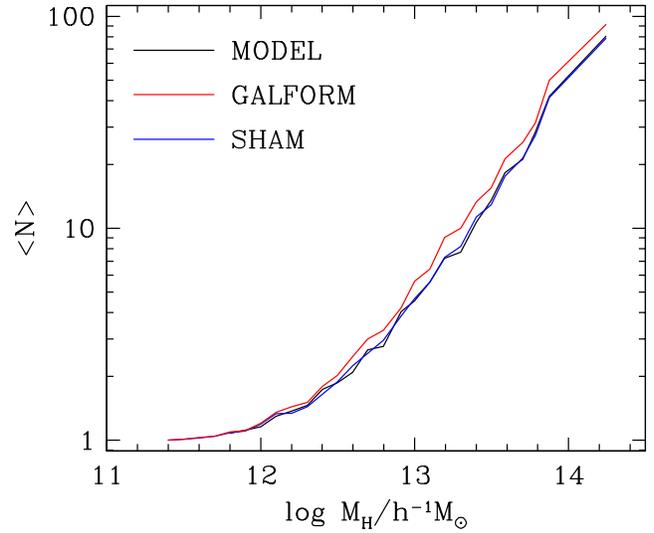}
}
\caption{
Halo Occupation Distribution of the model presented in this paper compared with two other models. GALFORM-GP14 is based on \protect\cite{gon14} and SHAM is abundance matching on the raw N-body simulation.
}
\label{fig:sub3}
\end{figure}

\begin{figure}
\centerline{
\epsfxsize=84mm
\epsfbox{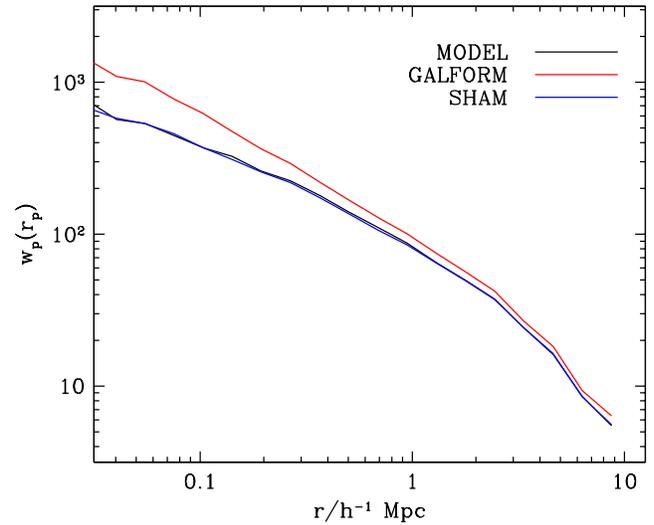}
}
\caption{
Projected two point correlation function of galaxies in the model presented in this paper compared with two other models. GALFORM-GP14 is based on \protect\cite{gon14} and SHAM is abundance matching on the raw N-body simulation.
}
\label{fig:sub4}
\end{figure}

\begin{figure}
\centerline{
\epsfxsize=84mm
\epsfbox{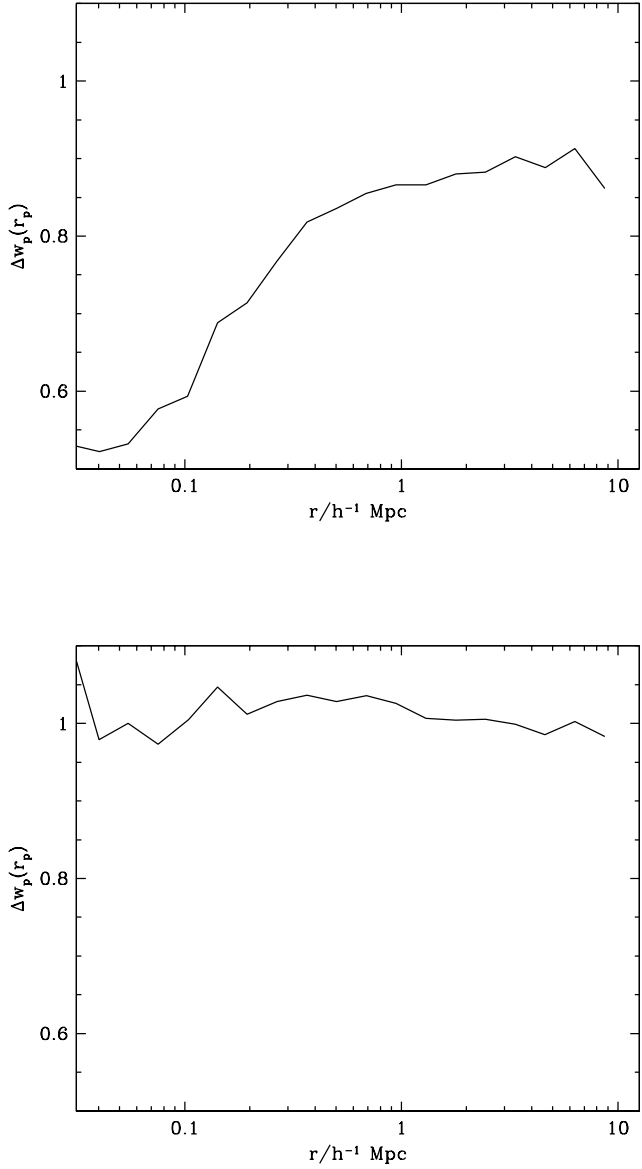}
}
\caption{
(Top) Ratio of the projected two point correlation function of galaxies in GALFORM-GP14 \protect\cite{gon14} and the model presented in this paper. (Bottom) Ratio of the projected two point correlation function of galaxies from abundance matching on the raw N-body simulation  and the model presented in this paper. 
}
\label{fig:sub5}
\end{figure}

Our model can be applied to generate mock galaxy catalogues from N-body simulations. Furthermore, it can also be applied to build mock galaxy catalogues from Monte-Carlo or other kinds of merger trees by drawing the energy and angular momentum of each subhalo from distributions similar to Fig. \ref{fig:sub6}, and then applying equation \ref{oureq} to determine the merger timescale.

\cite{campbell15} apply our model for subhalo evolution to the GALFORM semi-analytic model with galaxy stellar masses matched to observationally inferred stellar masses and find that it produces better agreement with the observed small scale clustering in SDSS at $z=0.1$ and GAMA at $z=0.2$ (see Figures 7 and 8 of \citealt{campbell15}). \cite{mcc16} apply our model to an N-body simulation that is similar to the Millennium Simulation, but with cosmological parameters determined by the PLANCK mission \citep{planck15}. They examine the halo occupation distributions and galaxy clustering and find better agreement with data from SDSS compared to GALFORM-GP14.
 
Our results are based on examining dark matter only simulations. Our model does not include the effect of baryons. Although stellar mass typically constitutes less than 10\% of the halo virial mass, baryons are more strongly concentrated than dark matter and more dense than dark matter at a given scale. As a result, they are more resistant to disruption. While including baryons reduces the likelihood of tidal disruption, it shortens the dynamical friction timescale. By comparing simulations with and without stellar bulges, \cite{bk09} find that the effect of baryons on merger time scales is typically less than 10\%. To shed further light on these issues, we plan to compare our prescription with hydrodynamic cosmological simulations in future work.

We emphasise that the model presented in this paper for the dynamical evolution of subhaloes uses the information in an N-body simulation, but can produce results that are not affected by artificial disruption of subhaloes due to limited resolution.

\section * {ACKNOWLEDGEMENTS}
I would like to thank John Helly for technical assistance during the course of this work and Spoorthy Raman for reading the manuscript and providing useful suggestions for improving the text. This work was supported by the Science and Technology Facilities Council [ST/L00075X/1].

\clearpage

\clearpage

\bibliographystyle{mn2e}

\end{document}